# Integration of Ether Unpacker into Ragpicker for Plugin-Based Malware Analysis and Identification

Erik Schaefer, Nhien-An Le-Khac, Mark Scanlon
School of Computer Science,
University College Dublin,
Belfield, Dublin 4, Ireland.

erik.schaefer@ucdconnect.ie, {an.lekhac, mark.scanlon}@ucd.ie

**Abstract:** Malware is a pervasive problem in both personal computing devices and distributed computing systems. Identification of malware variants and their families offers a great benefit in early detection resulting in a reduction of the analyses time needed. In order to classify malware, most of the current approaches are based on the analysis of the unpacked and unencrypted binaries. However, most of the unpacking solutions in the literature have a low unpacking rate. This results in a low contribution towards the identification of transferred code and re-used code. To develop a new malware analysis solution based on clusters of binary code sections, it is required to focus on increasing of the unpacking rate of malware samples to extend the underlying code database. In this paper, we present a new approach of analysing malware by integrating ETHER Unpacker into the plugin-based malware analysis tool, Ragpicker. We also evaluate our approach against real-world malware patterns.

## 1. Introduction

Malware has evolved from a pastime for hobbyist programmers and hackers into a powerful and effective money earner. Due to the increasing prevalence of technology in our lives, the opportunities of making money by exploiting these technologies' vulnerabilities rise proportionately. One problem is that malware developers have infinite degrees of freedom when it comes to the creation of a new piece of malware. Antivirus companies receive thousands of new malware samples daily. The time lag between the release of new malware threats and the response from the security industry allows the authors of such threats sufficient time to alter and/or improve their code to avoid detection. This statement is supported by the fact that in 2014 almost one million new malware threats were released every single day. Malware can result in substantial costs for business and governments. For example, the Zeus botnet alone is estimated to have caused over one hundred million USD of damages between 2007-2012 (Scanlon and Kechadi 2014) and malware attacks can cost companies over $6,500 per hour to rectify (Amin, 2016). The desire to avoid detection coupled with the often-lucrative nature of malware development means that there is a high probability that as new malware is developed it will likely utilise novel, unknown techniques. The discovery of these new techniques is essential to effectively defend against them.

Traditional malware detection relies greatly on file signatures, providing which is a reactive process of ending a malware, creating a signature (for example by hashing or extracting byte sequences) and pushing these signatures into filesystem scanners. However, this approach seems to fail against new malware code, which incorporates rootkit technologies and gets encoded to subvert anti-virus products. New malware detection approaches are based on behaviour analysis including static analysis and dynamic analysis. Static analysis extracts information about the binary code without executing the binary. On the other hand, dynamic malware analysis (DMA) extracts information about the code by observing what it does whilst it is running. These observations may include network communications (Scanlon and Kechadi, 2014), file and registry access (Sikorski and Honig, 2012) as well as modification, interaction with services and other behavioural activities (Wagener and Dulaunoy, 2008). Recently, a LE agency has developed an in-house malware analysis tool based on this DMA approach. However, this tool is based on ClamAV, a signature based unpacking with a low unpacking rate (only at 25%). In order to improve the performance of this tool, in this paper we present the integration of Ether (Dinaburg et al., 2008) generic unpacking capabilities into this malware analysis tool. Finally, we provide an evaluation of this new approach.

The rest of this paper is structured as follows: Section 2 shows the background of this research including the dynamic malware analysis. We present briefly the Ragpicker tool in Section 3. We describe our approach of implementing and integrating of Ether Unpacker in Section 4 and discuss on experiment results in Section 5. We conclude and discuss on future work in Section 6.

## 2. Dynamic Malware analysis

Most dynamic malware unpacking tools help to automate the unpacking process, try to create the unpacked code or the modified code by unpacking the stub on the runtime due to decompression or decryption. This section explores PolyUnpack (Royal et al., 2006), it firstly disassembles the packed file to build a static code model of the program. The program is separated into two sections, a set of sequences of code and a set of data. Then the program is single-step executed and the current instruction is compared with the code section of the model. The unpacked code is identified if it does not exist in the static model. Due to the use of disassembling and single stepping, this approach significantly increases computational overhead.

Renovo (Kang et al., 2007) monitors each instruction and tracks instructions that perform memory writes and control transfers. If it is a memory write instruction, the corresponding destination memory is marked as dirty which means it is newly generated. If any dirty memory is executed, one layer of unpacking is finished. The address pointed by the instruction pointer is identified as the OEP. This approach supports multiple layers of unpacking. However, like PolyUnpack, Renovo is an instruction-level approach and suffers from additional performance overhead.

Linke (2016) proposed a machine learning based method for both static and dynamic malware analysis. Despite the authors' method is efficient, it is however required unpacked and unencrypted malware patterns.

Safron (Danny Quist, 2007) uses a modified version of OllyBonE (Stewart, 2007) to dispatch information on the page-faults and run Intel PIN (Berkowits, 2012), an instrumentation tool, to monitor the flow of executables. If there is any execution control transfers to dynamically created or modified pages, a memory is dumped for further analysis. It is noted by the author of Safron that this approach is slow and does not cope with standard anti-debugging tricks (Ferrie, 2009) due to the use of PIN. OmniUnpack (Martignoni et al., 2007) tracks execution at the page level. The termination of unpacking is determined only when the overwritten pages is executed and the program is about to invoke a dangerous system call. At this time, a malware detection tool is used to check whether the unpacked file is a malware. If it is not, then the execution is resumed. Ether (Dinaburg et al., 2008) is a malware analysis framework which leverages hardware virtualisation extensions (specifically Intel VT) to remain transparent to malicious software. Ether is a generic malware analyser with the main target of detecting anti-detection techniques. The main advantages of Ether can be listed as follows:

1. It can detect self-modification (which may indicate unpacking). For binaries that are found to be possibly self-modifying, it can identify the code mechanism that carries out the unpacking and use this to unpack the binary without any prior knowledge about the packing algorithm used.
2. Ether's standard control and data flow analyses can be applied to this code modification mechanism to find (and possibly neutralise) dynamic defences, time/logic bombs, etc.

## 3. RAGPICKER

### 3.1 Integration point into the Analysis Process

For the purposes of the work presented as part of this paper, Ragpicker (a plugin based malware crawler) was chosen as the underlying malware analysis tool. Ragpicker currently has an embedded unpacking module based on ClamAV, which is used for comparison purposes. Ragpicker's analysis process chain starts with the crawling modules, as can be seen in Figure 1. Here the user can select malware feeds where the malware will be downloaded directly or an already known website with bad reputation will be parsed for malware binaries. Another option provided is to download entire malware databases, such as the onemalwr.com offers for free. Numerous crawling modules are included, but are not activated in the standard configuration.

After downloading the malware using the crawling modules, they will be saved in a MongoDB (Chodorow, 2013) database or software called VxCage (Guarnieri, 2015), representing a file based malware repository usable over a REST API. Using one or both options for repository purposes, Ragpicker now provides an enrichment of information for each sample gathered using the available processing modules. These modules provide basic analysis such as calculating the hash values, identifying the mime-type, collecting information from virustotal.com, etc. One of this processing modules checks whether the malware is packed using the `pefile/peutils` python modules by Ero Carrera (2014). In this function the sections of the PE file will be analysed. If enough sections look like they contain compressed data and the data makes up for more than 20% of the total file size then the function will return True. To determine that there must be compressed or encrypted data, the function uses algorithms to calculate the entropy of the relevant sections. The threshold value used to demonstrate that the section is packed or even not is 7.4. The value of 7.4 is empirical and determined by the community, based on looking at a few files packed by different packers.

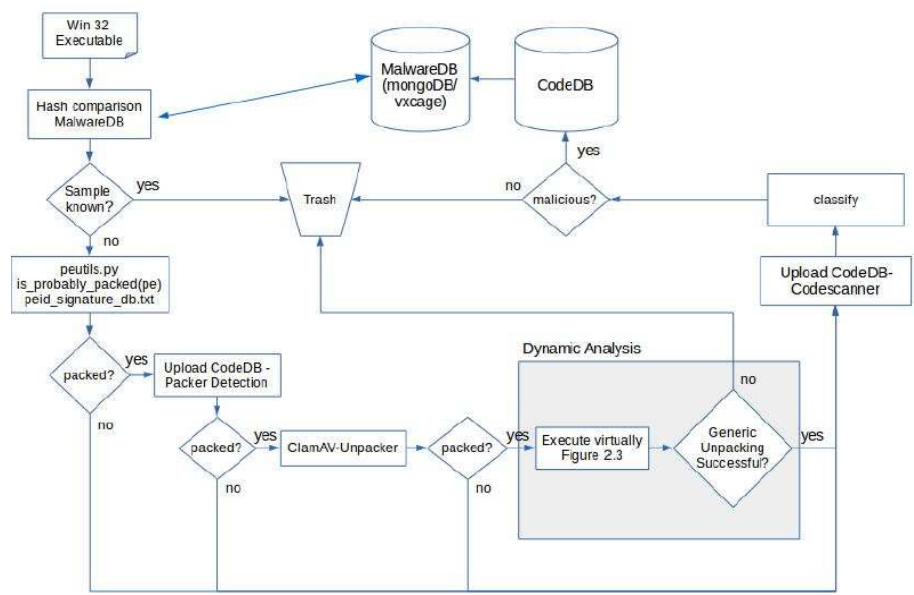

**Figure 1:** Unpacking Process Chain

As described by Zwanger et al. (2014) entropy is not the best indicator to detect a packed binary. High entropy occurs mostly in two places, in the code section and in the packed payload. Normally the code section has a smaller entropy than the packed payload, but that is unreliable as one will find packed payloads can also have low entropy. The Codescanner of the CodeDB operates inverted. First of all, it searches for sections and spaces that are definitely not entropic and suppresses them (Data, Strings, Null-Blocks, etc). Then only encrypted payload or the code section remains and can be observed with algorithms to detect the packer. An abstract description could be the proof of the existence of the relationship of 40% exposed code section, 10% data section and 50% packed payload. Every result shortening the 40% readable code section will mark the sample packed. This packer detection procedure seems to be more reliable than the way peutils.py does.

### 3.2 Evaluation of Ragpicker's ClamAV Unpacking Module

ClamAV is an open source (GPL) anti-virus engine used in a variety of situations including email scanning, web scanning, and end-point security. It provides several utilities including a flexible and scalable multi-threaded daemon, a command-line scanner and an advanced tool for automatic database updates. Using a special command-line option ClamAV is able to unpack different but already known packing algorithms like UPX, NSPACK, UPACK, ASPACK and some more. To utilise this functionality flexible a python function was written using the Linux command-line ClamAV-Tool clamscan. The unpacking rate of the clamscan binary was determined in two different ways. The first was the ability to download a malware sample-set daily in one package of hundred different samples from the commercial provider, (MalwareMonster, 2015), for free. The second way used the Ragpicker crawler again to put the final solution to the test. Using both methods, a static set and a dynamic continuously loading malware stream was used for evaluation. The results observing these different feeds are presented in Figure 2 and 3. Examples of utilised packers unpacked with ClamAV and identified with the `peid` signature database are:

- UPX - www.upx.sourceforge.net
- UPX v0.89.6 - v1.02 / v1.05 - v1.22 DLL
- Safeguard 1.03 - Simonzh
- Borland Delphi 3.0

At the time of writing, examples of packers that cannot be unpacked with ClamAV were:

- AHTeam EP Protector 0.3 (fake PCGuard 4.03-4.15) - FEUERRADER
- PECompact 2.0x Heuristic Mode – Jeremy Collake
- NsPack 2.9 - North Star
- Packman v1.0 - Brandon LaCombe

| Description | Day 1 | Day 2 | Day 3 | Day 4 | Day 5 |
|---|---|---|---|---|---|
| Counter off all Files: | 100 | 100 | 100 | 100 | 100 |
| Counter off PE32-Files: | 90 | 84 | 100 | 82 | 99 |
| $CodeDB$ identified Packed Files: | 36 | 61 | 64 | 41 | 72 |
| $ClamAV$ unpacked PE32-Files: | 14 | 1 | 5 | 1 | 1 |
| Unpacking Rate ($ClamAV/CodeDB$) in Percent: | 38.9% | 0.2% | 7.8% | 2.4% | 1.4% |

**Figure 2:** Unpacking Rates using ClamAV

| Description | Crawled Samples |
|---|---|
| Counter off all Files: | 18576 |
| Counter off PE32-Files: | 18576 |
| $CodeDB$ identified Packed Files: | 6963 |
| $ClamAV$ unpacked PE32-Files: | 1689 |
| Unpacking Rate in Percent: | 24.3% |

**Figure 3:** Ragpicker (using ClamAV) Unpacking Rate over 10 Week Testing Window

**4. Implementation and Integration of Ether Unpacker**

The Ether system was developed in 2008 by a group of scientists (ArtemDinaburg, 2008) and is a malware analysis framework which leverages hardware virtualisation extensions (specifically Intel VT) to remain transparent to malicious software. Ether is a generic malware analyser with the main target of detecting anti-techniques. Ether is transparent to malware and capable of handling most of the current anti-technique mechanisms. The limitations and threats that time were tolerable relative to others presented solutions.

**4.1 Preparation**

It is necessary to install the Ether machine from scratch to be able to integrate this approach into the parallel developed analysis cascade and to be able to start the evaluation regarding the quality and performance of the unpacking capability. The operation system Ether was built for was Debian 5 with a working XEN Hypervisor 3.1.0 implementation. The developers published the whole source code and a rudimentary documentation so that everyone who is interested could test the implied capabilities. As a prerequisite, the server hardware should have the Intel VT chip-set activated in BIOS. While reading through some papers that discussed Ether's pro and cons there where references to use the AMD VT-X chip-set to handle Ether detection or anti-analysing functions, but for the time working on the paper there was no hardware with an AMD chip-set available. That should be taken into account for the future work when the results of the Ether integration are worthwhile.

Next to the delivered installation description there were some FAQ and mailing list entries that explained how to avoid errors while installing the system. Several undocumented traps have to be circumvented. To get to know the main functionalities of Ether and how the system deals with malware samples, a virtualised WIN XP SP2 32-bit guest has been installed from scratch, too.

In this first Ether testing phase, it was immediately clear that Ether is only a prototype and not fully functional. The standard installation is not ready for the deployment as a generic unpacker for batch processing tasks that consumes one malware sample after another. The main focus lied on the controlled single-stepping execution of the sample which shall be analysed in one virtual machine. The great advantage in comparison to an agent based implementation like the Cuckoo Sandbox (Foundation, 2015) is that the machine can be monitored from the outside without any special manipulation of the guest that can be detected. Most of the anti-techniques, are eliminated that way. The control of XEN that time was not that comfortable like one can enjoy these days. For example, the very useful snapshot mechanism to recover a system state very fast does not exist. To prepare the machine for the purpose of this paper a higher automation capability is needed than this old XEN version facilitates out of the box to boost the analysis capacity.

**4.2 Software customisations**

The Ether source code only makes a prototype available where either one can follow the instructions of an executed binary inside a virtual machine one after another or an image of a beforehand provided process name will be dumped to a configured folder while detecting a memory write attempt. The execution of the potential malicious binary will not stop automatically. This prototype functionalities are implemented rudimentary and show several bugs that make the whole system unstable.

To control the XEN Hypervisor, some more functionalities were needed and have been configured and developed further on. Because the research phase of the paper was limited in time the organising features and costly investigated analysis time slots were implemented using bash scripts. The Ether unpacking function is not able to detect the end of the unpacking process. In the prototype, the executed binary stays in a loop waiting for a manual break though the unpacking concluded yet. As a matter of fact, dealing with different kind of malware and the Ether unpacking capabilities, one will find out that it is not possible to know exactly when the whole unpacking process is done. One will discover that malware could have several packing layers that cannot be predicted. Some research and evaluations were necessary to prepare expedient decisions for the configuration and control of the system in a most effective way. After the manual execution of different kind of packed binaries, it made sense to insert some interrupts or test loops. The above-mentioned snapshot mechanism, nowadays a standardised feature in virtual machine environments, has been recreated to optimise the automation capabilities that are needed for the integration into the analysis cascade later.

All in all, the use of not the newest available hardware and state-of-the-art technologies and especially the single-stepping approach of Ether means a long cycle for each malware sample that has to be unpacked. The system has been configured to unpack three samples simultaneously using three comparable virtual machines. One can state, working some weeks with that special configuration, the average analysis process of one samples lasts around 9 minutes. The boot process of each virtual machine was configured to 45 seconds, after measuring the time required not running into errors. One will experience that each booting process is different regarding the time it needs to get ready for the malware execution, though the virtual machine configuration is the same. Time influencing factors are the CPU load of the host system, the network availability and so on. In the virtual machine, all security mechanisms are deactivated. The samples are copied to the machine using SCP. The execution of the samples will be enabled per SSH command. When a memory write attempt is triggered with the Ether hypervisor listener then the relevant process will be dumped and saved outside the machine. The scripts then wait another 30 seconds to lead all unpacking functions finish. Additionally, Ether tries to fix the process memory dump to be executable afterwards. Therefore, it guesses the missing sections like the import address table (IAT). Working with several samples over some time showed that there is often more than one memory write attempt in that 30 seconds waiting loop. For every memory write attempt a separate process dump will be created and saved on the systems hard-disk. The Ether worker script stops automatically after 12 minutes' runtime when no unpacking shots of the software are detected. After an evaluation round, one can state that most of the samples were unpacked in that period of time. That helps to be resource effective regarding the duration of the execution of each sample and the unpacking process.

In the final configuration, used to solve the primary issue of a low ClamAV unpacking rate, the Ether machine was integrated listening permanently in one local directory for samples handed over and an output directory where the unpacked samples can be fetched. If Ether is not able to find the OEP after 12 minutes a text-file is created that informs about the failed state. The continuous run of the listener was ensured using a script which starts every minute controlled by a system scheduler.

In case the system crashes while analysing samples then after reboot the restart of the listener automatically cleans up the old aborted analysis. More on that system crashes will be stated in the section that describes the first impressions using Ether.

## 5. Evaluation of Ether Unpacker Integration

### 5.1 Assessments to the Runtime

All 101 samples were uploaded to the Ether unpacker. At the time of writing, the setup was configured to use three virtual machines in parallel. In the log-files of each sample unpacking attempt, the analysis runtime was determined and was found to be in the range outlined in Figure 4. Using the average time of 9 minutes 11 second, an estimate of the total duration for unpacking all samples is calculable. The overall unpacking duration was over 5 hours using the Ether unpacker in the parallel configuration.

| Ether Runtimes | Time Stamps |
|---|---|
| Fastest Runtime: | 4 min 43 sec |
| Slowest Runtime: | 12 min 29 sec |
| Average Runtime: | 9 min 11 sec |
| Complete Runtime (3 VMs): | 5 hours 20 min |

**Figure 4:** Ether Unpacker Runtime

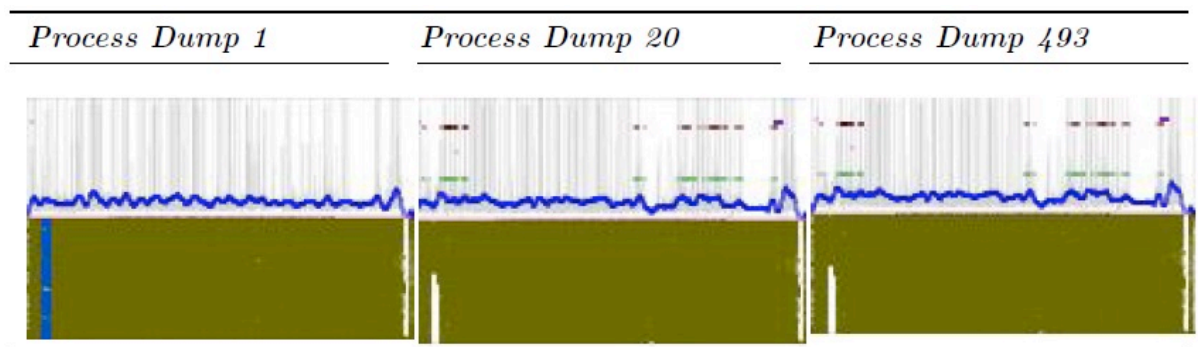

**Figure 5:** CodeDB Thumbnails Process Dumps of file-20150405-00097.exe

### 5.2 The Single Unpacked Sample

Every successful unpacking attempt creates a container file containing all relevant process dumps, an unpacking log file, and the original malware sample. The count of process dumps of one unpacked sample was found to differ from 1 to 493 dumps for the set of samples tested. Figure 5 shows that only the first dump contains visible code. The other dumps until the last one also contained no visible code. Therefore, the decision was made that only the first dump shall be used for further analysis steps and especially for the automatic onward transfer to the CodeDB.

### 5.3 Ether Unpacking Summary

The results after executing all 101 packed samples in Ether's virtual machine configuration were promising. 61 prosperous unpacking attempts were counted. These results mean an average unpacking rate of 60.3%. Comparing against ClamAV's unpacking rates (outlined in Figure 2), Ether results in a significant improvement in the unpacking rates, as outlined in Figure 6. The error adjusted packed files (APF) are the files detected packed by the CodeDB initially, reduced by the successful ClamAV unpacking attempts and diminished again removing any corrupt files.

### 5.4 The CodeDB Integration

A python script was developed to extract the first process dump of the tar-file and uploads it automatically to the CodeDB. There, once again, a packer check is one of the first criteria the CodeDB executes. The code database needs more then 40% code in relation to the packed payload. After evaluating all successfully unpacked samples in combination with the CodeDB, only 5 samples were accepted into the database unpacked. All other samples were rejected with the status message "\packed". In general, this does not mean that the dumps do not contain any unpacked sections merely not passing the required 40% threshold. The unaccepted samples can still be useful for other purposes, for example the examination of the embedded and now in-clear text strings. The requirements of the CodeDB demand more quality regarding the achieved unpacking results.

| Description/Day | Day 1 | Day 2 | Day 3 | Day 4 | Day 5 |
|---|---|---|---|---|---|
| $CodeDB$ identified Packed Files: | 36 | 61 | 64 | 41 | 72 |
| $ClamAV$ Unpacked PE32-Files: | 14 | 1 | 5 | 1 | 1 |
| Unpacking Rate ($ClamAV/CodeDB$) in Percent: | 38.9% | 0.2% | 7.8% | 2.4% | 1.4% |
| Error Adjusted Packed Files (APF): | 14 | 19 | 24 | 19 | 25 |
| Unpacked Files: | 11 | 15 | 15 | 6 | 15 |
| Unpacking Rate ($Ether$/APF) in Percent: | 78.6% | 78.9% | 62.5% | 31.6% | 60% |

**Figure 6:** Unpacking Rates using the Ether Unpacker

### 6. Conclusion and Future work

In this paper, the results from the integration and evaluation of the Ether generic unpacking tool into Ragpicker were presented. Unpacked binaries are necessary to be able to cluster the analysis results on the code sections and reduce the crawled samples by assigning them to code families. The work presented can also contribute to answering several research questions, such as: (i) Does the quality of the unpacked code sections match the requirements? (ii) Are the selected systems able to overcome the race against the established and continuously enhanced anti-techniques? The work presented as part of this paper still leaves significant room for further research and improvement. Even though the simple aim to enhance the unpacking rate was achieved, dynamic malware analysis systems have much room for improvement. The contribution of this paper can assist the investigators in forensics in industrial control systems (Vliet, 2016) or in forensic acquisition and analysis of VoIP applications (Sgaras, 2015) where the packed and encrypted malware patterns are also the challenges or even in triage process for front-line forensic personnel (Hitchcock et al. 2016).

### References

Amin, M. (2016). A Survey of Financial Losses Due to Malware. In Proceedings of the Second International Conference on Information and Communication Technology for Competitive Strategies (p. 145). ACM.

Berkowits, S. (2012). Pin - a dynamic binary instrumentation tool. https://software.intel.com/en-us/articles/pin-a-dynamic-binary-instrumentation-tool.

Carrera, E. (2014, 10). pefile/peutils - https://github.com/erocarrera/pefile.

Chodorow, K. (2013) MongoDB: The Definitive Guide, 2nd Edition Powerful and Scalable Data Storage. O'Reilly Press May 2013

Danny Quist, V. (2007). Covert debugging circumventing software armouring techniques. Black hat briefings USA.

Dinaburg, A., Royal, P., Sharif, M., & Lee, W. (2008). Ether: Malware analysis via hardware virtualization extensions. In (p. 12). (Georgia Institute of Technology, Atlanta, GA, USA)

Ferrie, P. (2009). Anti unpacker tricks 2. Virus Bulletin.

Guarnieri, C. (2015). Vxcage. https://github.com/botherder/vxcage

Hitchcock, B., Le-Khac, N-A. and Scanlon, (2016) Tiered Forensic Methodology Model for Digital Field Triage by Non-Digital Evidence Specialists, *Digital Investigation* Vol. 16(13S), p.S75-S85, 2016

Kang, M. G., Poosankam, P., & Yin, H. (2007). Renovo: A hidden code extractor for packed executables. In (p. 8). Alexandria, Virginia, USA. (Carnegie Mellon University)

Linke, A., Le-Khac, N-A. (2016) Control Flow Change in Assembly as a Classifier in Malware Analysis, 4th IEEE International Symposium on Digital Forensics and Security, Little Rock, AR, USA, April 2016.

MalwareMonster. (2015). Malware sample data feed - http://www.amanatsumikann.com/download/.

Martignoni, L., Christodorescu, M., & Jha, S. (2007). Omniunpack: Fast, generic, and safe unpacking of malware. , 10. (Universit degli Studi di Milano, IBM Research, University of Wisconsin)

Royal, P., Halpin, M., Dagon, D., Edmonds, R., & Lee, W. (2006). Polyunpack: Automating the hidden-code extraction of unpack-executing malware. In (p. 10). (College of Computing Georgia Institute of Technology)

Sgaras, C., Kechadi, M-T., Le-Khac, N-A. (2015) Forensics Acquisition and Analysis of Instant Messaging and VoIP Applications, Computational Forensics, Springer Verlag, LNCS Vol. 8915, p.188-199, 2015

Scanlon, M. and Kechadi, M.T., (2013). Universal peer-to-peer network investigation framework. In Availability, Reliability and Security (ARES), 2013 Eighth International Conference on (pp. 694-700). IEEE.

Sikorski, M. and Honig, A. (2012). Practical malware analysis: the hands-on guide to dissecting malicious software. no starch press.

Stewart, J. (2007). Ollybone. http://www.joestewart.org/ollybone/.

Wagener, G. and Dulaunoy, A., (2008). Malware behaviour analysis. Journal in Computer Virology, 4(4), pp.279-287.

Zwanger, V., Gerhards-Padilla, E., & Meier, M. (2014). Codescanner: Detecting (hidden) x86/x64 code in arbitrary files.. (Institute of Computer Science 4, University of Bonn)

Foundation, C. (2015). Cuckoo sandbox. http://www.cuckoosandbox.org/.

ArtemDinaburg, Paul Royal,M. S.W. L. (2008). Ether: Malware analysis via hardware virtualization extensions. page 12. CCS'08. Georgia Institute of Technology, Atlanta, GA, USA.

Vliet, P-V., Kechadi, M-T., Le-Khac, N-A. (2015) Forensics in Industrial Control System: A Case Study, Springer Verlag, LNCS Vol. 9588, p.147-156, 2015